\documentclass[fleqn,twoside,twocolumn,nofootinbib,showkeys]{revtex4} 
\usepackage[sec,nocpr]{ujp} 

\begin{document}
\title[Asymmetric Diamond Ising--Hubbard Chain with
Attraction]%
{ASYMMETRIC DIAMOND ISING--HUBBARD\\ CHAIN WITH ATTRACTION}%
\author{B.M. Lisnyi}
\affiliation{Institute for Condensed Matter Physics, Nat. Acad. of
Sci. of
Ukraine}%
\address{1, Svientsitskii Str., Lviv 79011, Ukraine}%
\email{lisnyj@icmp.lviv.ua}
\udk{538.953, 538.955} \pacs{75.10.Pq, 75.40.Cx,\\[-3pt] 75.10.Jm}
\razd{\secix}

\setcounter{page}{195}%
\autorcol{B.M.\hspace*{0.7mm}Lisnyi}

\begin{abstract}
The ground state and thermodynamic properties of an asymmetric
diamond Ising--Hubbard chain with the on-site electron-electron
attraction has been considered. The problem can be solved exactly
using the decoration-iteration transformation. In the case of the
antiferromagnetic Ising interaction, the influence of this
attraction on the ground state and the temperature dependences of
the magnetization, magnetic susceptibility, and specific heat has
been studied.
\end{abstract}

 \maketitle

\section{Introduction}

The spin-chain problem, which is solved exactly with the use of the
decoration-iteration transformation~\cite{s5,s6,spla10,rojas11}
attracts interest, because it allows certain features in the
properties of complicated spin systems and magnetic materials to be
studied~\cite{ki05l,ki05ptp,s3}. These are the intermediate plateaux
on magnetization curves and extra maxima on the temperature
dependence of the heat capacity. The spin chains also enable the
interrelation between a geometrical frustration and quantum
fluctuations to be analyzed \cite{dos2,scmp09,lis3}. Moreover, they
can serve as models for the quantitative description of magnetic
properties of materials \cite{s3,exp10}. That is why the
one-dimensional models, which can be solved exactly with the use of
the decoration-iteration transformation, are actively studied
\cite{oh03,dos1,valpa08,valjpcm08,dos3,dos4,ohcmp09,ohprb09,rpla11,rprb11}.

Work \cite{dos3}, in which the properties of an asymmetric diamond
Ising--Hubbard chain were considered without refard for the
on-site electron-electron interaction, has started the researches
of exactly solved (by applying the decoration-iteration
transformation) Ising--Hubbard systems \cite{sprb09,gjpcm11}. This
chain reveals the following features: the 0 and 1/3 magnetization
plateaux \cite{dos3}, one \cite{dos3} or two \cite{lis2}
additional low-temperature peaks in the zero-field heat capacity,
and the considerable adiabatic magnetocaloric coefficient
\cite{dos4}. The influence of the on-site Coulomb repulsion of
electrons on the property of this chain was studied in work
\cite{lis2}. In particular, in the interval with substantial
repulsion, the zero-field heat capacity was shown to have an
additional maximum at high temperatures.

If the chain of electrons interacts with local phonons, an effective
on-site electron-electron attraction can be realized in it
\cite{prb85,zetf96,pla03}. By analogy, in this work, the properties
of an asymmetric diamond Ising--Hubbard chain \cite{dos3} with the
on-site electron-electron attraction is considered. In particular,
in the case of the Ising antiferromagnetic interaction, when
a geometrical frustration in the chain emerges, the influence of
this attraction on the ground state and thermodynamic properties
will be analyzed.

\section{Model and Its Exact Solution}

Consider an asymmetric diamond Ising--Hubbard chain \cite{dos3,lis2}
with the on-site electron-electron attraction (Fig.~1) in a magnetic
field. The primitive cell of this chain is determined by the $k$-th
and $(k+1)$-th nodes, both occupied by the so-called Ising spins,
$\mu_{k}$, which are coupled with neighbor spins by means of the Ising
interaction. Two electrons with the on-site attraction between them
execute quantum jumps over two interstitial positions $(k,1)$ and
$(k,2)$ in the primitive cell.

\begin{figure}
\includegraphics[width=7.0cm]{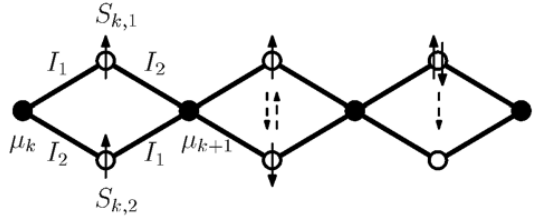}
\vskip-3mm\caption{Schematic diagram of a fragment in an asymmetric
diamond Ising--Hubbard chain. Solid and hollow circles denote nodes
and interstitial positions, respectively. For the $k$-th primitive
cell, the nodal spins, $\mu_{k}$, the $z$-components of total spins
in the interstitial positions, $S_{k,i}$, and the parameters of the
Ising interaction for bonds along the diamond sides,
$I_{i}%
$, are indicated  }
\end{figure}

The Hamiltonian $\mathcal{H}$ of the chain is written down as a sum of cell
Hamiltonians $\mathcal{H}_{k}$,
\[
\mathcal{H}=\!\!\sum\limits_{k=1}^{N}\mathcal{H}_{k},
\]\vspace*{-5mm}
\[
\mathcal{H}_{k}{=}\!\!\sum\limits_{\sigma{\in}\{\uparrow,\downarrow
\}}\!\!\!t\big(c_{k,1;\sigma}^{\dag}c_{k,2;\sigma}+c_{k,2;\sigma}^{\dag
}c_{k,1;\sigma}\big)\!-\]\vspace*{-5mm}
\[-\!\sum\limits_{i=1}^{2}Un_{k,i;\uparrow}n_{k,i;\downarrow
}+
\]\vspace*{-5mm}
\[
{}+\mu_{k}(I_{1}S_{k,1}+I_{2}S_{k,2})+\mu_{k+1}(I_{2}S_{k,1}+I_{1}S_{k,2})-
\]\vspace*{-7mm}
\begin{equation}
{}-h_{\text{e}}(S_{k,1}+S_{k,2})-\frac{1}{2}{h_{\text{i}}}(\mu_{k}+\mu_{k
+1}),
\label{Hk}%
\end{equation}
where $N$ is the number of primitive cells in the chain;
$c_{k,i;\sigma}%
^{\dag}$ and $c_{k,i;\sigma}$ are the operators of creation and
annihilation,
respectively, of an electron with the spin
$\sigma\in\{\uparrow,\downarrow\}$
at the interstitial position $(k,i)$;
$n_{k,i;\sigma}=c_{k,i;\sigma}^{\dag
}c_{k,i;\sigma}$ is the operator of electron number; $\mu_{k}$ is the
$z$-component of the spin-1/2 operator; $S_{k,i}=(n_{k,i;\uparrow}%
-n_{k,i;\downarrow})/2$ is the $z$-component of the operator of
total electron spin at the interstitial position $(k,i)$; $t$ is the
transfer integral; $U$ is the magnitude of the on-site
electron-electron attraction ($U\geq0$); $I_{1}$ and $I_{2}$ are the
parameters of the Ising interaction for the bonds along the diamond
sides (this interaction being identical only for collinear bonds)
(Fig.~1); and $h_{\text{e}}$ and $h_{\text{i}}$ are magnetic fields
that act on electron spins and Ising spins, respectively. It should
be noted that Hamiltonian (\ref{Hk}) also corresponds to a simple
Ising--Hubbard chain (the nodes and interstitial positions are
aligned), in which the Ising spin $\mu_{k}$ is coupled with the
first, $I_{1}$, and second, $I_{2}$, neighbors.

The replacement of the on-site electron-electron attraction  by a
repulsion in Hamiltonian (\ref{Hk}) transforms it into the chain
Hamiltonian from work \cite{lis2}, which was solved exactly by
applying the decoration-iteration transformation. Therefore, the
exact solution for our chain can be obtained by replacing the
on-site electron-electron repulsion by the attraction
($U\rightarrow-U$) in all results of work \cite{lis2}. The spectrum
of the Hamiltonian $\mathcal{H}_{k}$ obtained in such a way
\mbox{looks like}
\[
\mathcal{E}_1 (\mu_{k},\mu_{k+1}) = \frac{I_1 + I_2}{2}(\mu_{k} +
\mu_{k+1}) \!-\!h_{\textrm{e}} -\]\vspace*{-5mm}
\[-\frac{h_{\textrm{i}}}{2} (\mu_{k} \!+\!\mu_{k+1}),
\]\vspace*{-5mm}
\[
\mathcal{E}_2 (\mu_{k},\mu_{k+1}) = -\frac{I_1 + I_2}{2}(\mu_{k} +
\mu_{k+1})\! +\!h_{\textrm{e}} -\]\vspace*{-5mm}
\[-\frac{h_{\textrm{i}}}{2} (\mu_{k}
\!+\!\mu_{k+1}),
\]\vspace*{-5mm}
\[
\mathcal{E}_3 (\mu_{k},\mu_{k+1}) = \Lambda_1 |\mu_{k} -\mu_{k+1}|
-\frac{h_{\textrm{i}}}{2} (\mu_{k} +\mu_{k+1}),
\]\vspace*{-5mm}
\[
\mathcal{E}_4 (\mu_{k},\mu_{k+1}) =  \frac12\left(\!\sqrt{U^2 +
16t^2} -U\!\right)|\mu_{k} +\mu_{k+1}| +
\]\vspace*{-5mm}
\[
{} + \Lambda_2 |\mu_{k} -\mu_{k+1}| -\frac{h_{\textrm{i}}}{2}
(\mu_{k} +\mu_{k+1}),
\]\vspace*{-5mm}
\[
\mathcal{E}_5 (\mu_{k},\mu_{k+1}) = -\frac12\left(\!\sqrt{U^2 +
16t^2} +U\!\right)|\mu_{k} +\mu_{k+1}| +
\]\vspace*{-5mm}
\[
{} + \Lambda_3 |\mu_{k} -\mu_{k+1}| -\frac{h_{\textrm{i}}}{2}
(\mu_{k} +\mu_{k+1}),
\]\vspace*{-5mm}
\begin{equation}
\mathcal{E}_6 (\mu_{k},\mu_{k+1}) = -U -\frac{h_{\textrm{i}}}{2}
(\mu_{k} +\mu_{k+1}), \label{Ek}
\end{equation}
where $\Lambda_{3}<\Lambda_{1}<\Lambda_{2}$ are the eigenvalues of the
matrix
\[
\mathcal{L}=\left(\!\!
\begin{array}
[c]{ccc}%
0 & \frac{I_{1}-I_{2}}{2} & 0\\
\frac{I_{1}-I_{2}}{2} & 0 & 2t\\
0 & 2t & -U
\end{array}
\!\!\right)\!  .
\]
From this spectrum, the ground state and the parameters of
a decoration-iteration transformation can be determined \cite{lis2}.

\section{Numerical Results and Their Discussion}

Consider the properties of our chain in the case of the antiferromagnetic
Ising
interaction ($I_{i}>0$), when there exists a geometrical frustration in
it.
The magnetic fields are put to be identical,
$h=h_{\text{i}}=h_{\text{e}}$.
Without any loss of generality, we adopt that $I_{1}\geq I_{2}$ and introduce
the
parameter $\Delta I=I_{1}-I_{2}$, similar to what was done in work
\cite{dos3}. Let us pass to the dimensionless parameters,
\[
\tilde{t}=\frac{t}{I_{1}},\quad\tilde{U}=\frac{U}{I_{1}},\quad\Delta\tilde
{I}=\frac{\Delta I}{I_{1}},\quad\tilde{h}=\frac{h}{I_{1}}.
\]
It will be recalled that the parameter $\Delta\tilde{I}\in\lbrack0,1]$
characterizes the degree of asymmetry for the Ising interaction along the
diamond
sides \cite{lis2}.

Consider firstly the properties of the system in the ground state.
The latter corresponds to the minimum energy in spectrum (\ref{Ek})
for all possible configurations $\mu_{k}$ and $\mu_{k+1}$. Depending
on the parameters $\tilde{t}$, $\tilde{U}$, $\Delta\tilde{I}$, and
$\tilde{h}$, the system concerned can be in four ground states~--
similarly to what take place at $\tilde{U}\leqslant0$
\cite{dos3,lis2}~-- namely, the saturated paramagnetic (SPA),
ferrimagnetic (FRI), nonsaturated paramagnetic (UPA), and nodal
antiferromagnetic (NAF) states. The energies of those states per
cell are
\cite{lis2}%
\[
\tilde{\mathcal{E}}_{\text{SPA}}=1-\frac{\Delta\tilde{I}}{2}-\frac{3\tilde{h}%
}{2},\quad\tilde{\mathcal{E}}_{\text{FRI}}=-1+\frac{\Delta\tilde
{I}}%
{2}-\frac{\tilde{h}}{2},
\]\vspace*{-5mm}
\[
\tilde{\mathcal{E}}_{\text{UPA}}=-\frac{1}{2}\left(
\tilde{U}+\sqrt{\tilde {U}^{2}+16\tilde{t}^{2}}\right)
-\frac{\tilde{h}}{2},\quad\tilde
{\mathcal{E}}_{\text{NAF}}=\tilde{\Lambda}_{3},
\]

\begin{figure}%
\includegraphics[width=7.0cm]{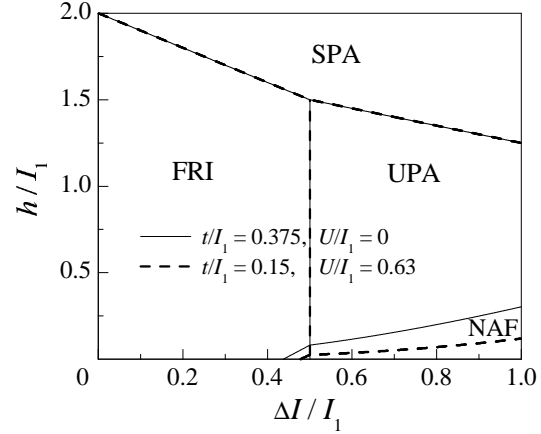}
\vskip-3mm\caption{Phase diagram $(\Delta\tilde{I},\tilde{h})$ for
the ground state. Transition lines for two sets of $\tilde{t}$- and
$\tilde{U}$-values are shown. The results obtained for $\tilde{U}=0$
coincide with those of work \cite{dos3}  }\vskip2mm
\end{figure}

\noindent
where $\tilde{\Lambda}_{i}$ are the eigenvalues of the matrix
$\tilde{\mathcal{L}%
}=\mathcal{L}/{I_{1}}$. The wave functions of those states are
\cite{lis2}
\[
|\text{SPA}\rangle=\prod\limits_{k=1}^{N}{|+\rangle}_{k}~{|\uparrow
,\uparrow\rangle}_{k,1;k,2},
\]\vspace*{-5mm}
\[
|\text{FRI}\rangle=\prod\limits_{k=1}^{N}{|-\rangle}_{k}~{|\uparrow
,\uparrow\rangle}_{k,1;k,2},
\]\vspace*{-5mm}
\[
|\text{UPA}\rangle=\prod\limits_{k=1}^{N}{|+\rangle}_{k}~\left[
\Psi_{5}%
^{++}\right]  _{k,1;k,2},
\]\vspace*{-5mm}
\[
|\text{NAF}\rangle=\left\{  \!\!
\begin{array}
[c]{l}%
\prod\limits_{k=1}^{N}\left\vert (-)^{k}\right\rangle _{k}~\left[
\Psi
_{5}^{(-)^{k}(-)^{k+1}}\right]  _{k,1;k,2}\\[4mm]%
\prod\limits_{k=1}^{N}\left\vert (-)^{k+1}\right\rangle _{k}~\left[
\Psi _{5}^{(-)^{k+1}(-)^{k+2}}\right]  _{k,1;k,2}
\end{array}
\right.  \!\!,
\]
where $|+\rangle_{k}=|\uparrow\rangle_{k},$ and
$|-\rangle_{k}=|\downarrow \rangle_{k}$ describe the state of spin
$\mu_{k}$, $|\uparrow,\uparrow
\rangle_{k,1;k,2}=c_{k,1;\uparrow}^{\dag}c_{k,2;\uparrow}^{\dag}|0\rangle$,
and the notation $(-)^{n}$ means the sign of $(-1)^{n}$. The other
notations are
\[
\Psi_{5}^{++}\!=\!A_{5}^{++}\big(|\uparrow,\downarrow\rangle+|\downarrow
,\uparrow\rangle\big)+B_{5}^{++}\big(|\uparrow\downarrow,0\rangle
+|0,\uparrow\downarrow\rangle\big),
\]\vspace*{-5mm}
\[
\Psi_{5}^{\pm\mp}\!=\!A_{5}^{\pm\mp}|\uparrow,\downarrow\rangle+A_{5}^{\mp\pm
}|\downarrow,\uparrow\rangle+B_{5}^{+-}\big(|\uparrow\downarrow,0\rangle
+|0,\uparrow\downarrow\rangle\big),
\]
where
\[
|\uparrow,\downarrow\rangle=c_{k,1;\uparrow}^{\dag}c_{k,2;\downarrow}^{
\dag
}|0\rangle,~~~~|\downarrow,\uparrow\rangle=-c_{k,1;\downarrow}^{\dag
}c_{k,2;\uparrow}^{\dag}|0\rangle,
\]\vspace*{-5mm}
\[
|\uparrow\downarrow,0\rangle=c_{k,1;\uparrow}^{\dag}c_{k,1;\downarrow}^
{\dag
}|0\rangle,~~~|0,\uparrow\downarrow\rangle=c_{k,2;\uparrow}^{\dag
}c_{k,2;\downarrow}^{\dag}|0\rangle,
\]
and the quantities $A_{5}^{++}$, $B_{5}^{++}$, $A_{5}^{+-}$,
$A_{5}^{-+}$, and
$B_{5}^{+-}$ are obtained from the corresponding coefficients in work
\cite{lis2} by replacing $\tilde{U}\rightarrow-\tilde{U}$.

\begin{figure}
\includegraphics[width=7.2cm]{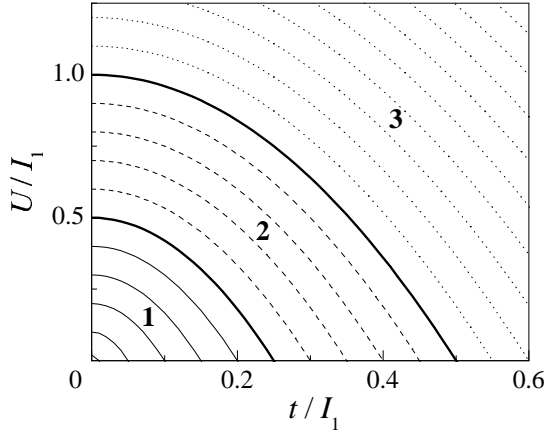}
\vskip-3mm\caption{Topological diagram $(\tilde{t},\tilde{U})$ for
the phase diagram of the ground state, $(\Delta\tilde{I},\tilde{h})$,
with \textquotedblleft equitopological\textquotedblright\ lines.
Bold lines separate the ranges of three typical topologies
designated by the corresponding figures  }
\end{figure}

\begin{figure}%
\includegraphics[width=7.5cm]{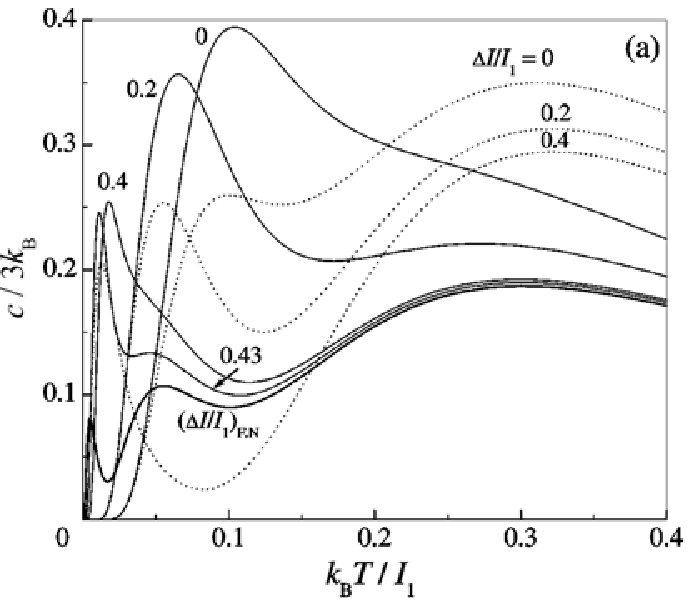}\\[2mm]
\includegraphics[width=7.5cm]{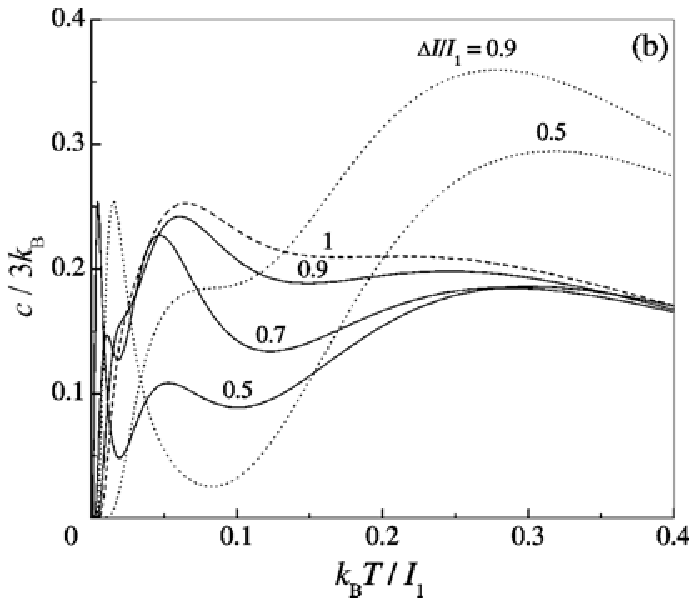}
\vskip-3mm \caption{Temperature dependences of the zero-field heat
capacity at various $\Delta\tilde{I}$:
(\textit{a})~$\Delta\tilde{I}\leq\Delta\tilde {I}_{\text{F.N}}$ and
(\textit{b})~$\Delta\tilde{I}>\Delta\tilde {I}_{\text{F.N}}$. Solid
and dashed curves correspond to the results obtained for
$\tilde{t}=0.15$ and $\tilde{U}=0.63$. Dotted curves show the
results obtained for $\tilde{t}=0.375$ and $\tilde{U}=0$, which were
exhibited in work \cite{dos3} except for the curve for
$\Delta\tilde{I}=0.4$ }\vskip1.5mm
\end{figure}

Let us consider the phase diagram for the ground
state,
$(\Delta\tilde{I},\tilde{h})$, which depends on the parameters
$\tilde{t}$ and
$\tilde{U}$. One of the possible phase diagrams
$(\Delta\tilde{I},\tilde{h})$ is
depicted in Fig.~2. The FRI and UPA states in it are separated by the
transition line $\Delta\tilde{I}_{\text{F%
$\vert$%
U}}=\tilde{\mathcal{T}}(\tilde{t},\tilde{U})$, where
$\tilde{\mathcal{T}%
}(\tilde{t},\tilde{U})=2-\tilde{U}-\sqrt{{\tilde{U}}^{2}+16\tilde{t}^{2
}}$. The critical point between the FRI and NAF states in the zero
field, $\Delta\tilde{I}_{\text{F.N}}$, is determined from the
equation $\Delta
\tilde{I}_{\text{F.N}}=2+2\tilde{\mathcal{E}}_{\text{NAF}}$.

The topology of the phase diagram $(\Delta\tilde{I},\tilde{h})$ for the
ground
state belongs to one of three types, depending on the parameters
$\tilde{t}$
and $\tilde{U}$. The topology type is determined by the parameter
$\tilde{\mathcal{T}}(\tilde{t},\tilde{U})$. The first type of the phase
diagram
topology is realized at $1\leq\tilde{\mathcal{T}}$. In this case, it
looks as
a part of Fig.~2 within the interval $[0,\Delta\tilde{I}_{\text{F%
$\vert$%
U}}]$ \cite{dos3,lis2}. The second topology type is realized at
$0<\tilde
{\mathcal{T}}<1$ (see Fig.~2). The third topology type is realized at
$\tilde{\mathcal{T}}\leq0$. In this case, the phase diagram
$(\Delta\tilde
{I},\tilde{h})$ looks like a part of Fig.~2 within the interval $[\Delta
\tilde{I}_{\text{F%
$\vert$%
U}},1]$, but now the line of the NAF $\leftrightarrow$ UPA transition always
begins at the point (0,0) \cite{dos3,lis2}.

It is convenient to represent the dependence of the $(\Delta\tilde{I}%
,\tilde{h})$ phase diagram topology on the parameters $\tilde{t}$ and
$\tilde{U}$ in the form of a topological diagram ($\tilde{t},\tilde{U}$),
which
is shown in Fig.~3. \textquotedblleft Equitopological\textquotedblright\
lines
in this diagram are described by the equation
$\tilde{\mathcal{T}}(\tilde
{t},\tilde{U})=$const. The changes of $\tilde{U}$ and $\tilde{t}$
along the
\textquotedblleft equitopological\textquotedblright\ line are reflected
in the
phase diagram $(\Delta\tilde{I},\tilde{h})$ as a displacement of only
those
lines that bound the range of the ground NAF state, which is demonstrated in
Fig.~2. In particular, as the parameter $\tilde{U}$ grows, the range of
the ground
NAF state in it decreases.

Now consider the influence of the on-site electron-electron attraction
on the
magnetization, magnetic susceptibility, and heat capacity in the regime
$\tilde{\mathcal{T}}(\tilde{t},\tilde{U})=\mathrm{const}$, i.e. along
an
\textquotedblleft equitopological\textquotedblright\ line. Numerical
calculations of those characteristics were carried out for a number of
points
$(\tilde{t},\tilde{U})$ along the \textquotedblleft
equitopological\textquotedblright\ line that belongs to region~2 in
Fig.~3 and
passes through the points $(0.375,0)$ and $(0.15,0.63)$, for which the
phase
diagram $(\Delta\tilde{I},\tilde{h})$ of the ground state is exhibited
in Fig.~2.

A comparison of the results obtained for the magnetization and the
magnetic
susceptibility at various points on the \textquotedblleft
equitopological\textquotedblright\ line shows that the field and
temperature
curves of magnetization and the temperature curve of magnetic
susceptibility
in the zero field shift at strengthening the attraction, similarly to what
takes
place at weakening the on-site electron-electron repulsion
\cite{lis2}. In
particular, if the attraction becomes stronger, the temperature curves of
total
and electron magnetizations shift downward, and the temperature curve
of
magnetization for Ising spins shifts upward.

The modification of the temperature dependence of the zero-field heat
capacity
under the influence of the attraction is shown in Fig.~4. In the absence of
attraction ($\tilde{U}=0$), the temperature curve of the heat capacity has
the main
and low-temperature maxima in a wide range of $\Delta\tilde{I}$
\cite{dos3}. In a certain vicinity of the critical point $\Delta\tilde
{I}_{\text{F.N}}$, it has another low-temperature maximum, which is the
closest to the zero temperature \cite{lis2}. While the attraction grows to a
definite value of about 0.42, the main maximum shifts toward lower
temperatures. However, if the attraction grows further, the maximum shifts
back
toward higher temperatures. As a result, the main maximum of the curves
that
correspond to small $\Delta\tilde{I}$ can merge with the low-temperature
maximum (Fig.~4). Moreover, starting from $\tilde{U}\simeq0.5$, there
emerges
an additional maximum in the temperature dependence of the heat capacity,
which is
located between the low-temperature and main maxima (Fig.~4). At
first,
this additional maximum exists within a small interval for
$\Delta\tilde{I}$
in the range of the ground NAF state. As the attraction becomes stronger, this
interval broadens and surrounds the critical point $\Delta\tilde
{I}_{\text{F.N}}$ (Fig.~4). The modification in the temperature
dependence of
the zero-field heat capacity induced by the attraction growth is associated
with
changes in spectrum (\ref{Ek}) of the cell Hamiltonian
$\mathcal{H}_{k}$.
Namely, as the parameter $\tilde{U}$ increases, the energies
$\mathcal{E}_{6}%
$, $\mathcal{E}_{4}({\Lambda}_{2})$, and
$\mathcal{E}_{3}({\Lambda}_{1})$ in
this spectrum decrease.

It is also worth noting that the temperature curve of the heat
capacity for $\tilde{U}=0.63$ has an additional low-temperature
maximum in a very close--much narrower than at
$\tilde{U}=0$--vicinity to the critical point
$\Delta\tilde{I}_{\text{F.N}}$. This maximum, which is the nearest
to the zero temperature, was described in detail in work
\cite{lis2}.\vspace*{-2mm}

\section{Conclusions}

In this work, the properties of an asymmetric diamond Ising--Hubbard
chain with the on-site electron-electron attraction are studied in the ground
state at finite temperatures. It is an example of
the exact solution obtained with the use of the decoration-iteration
transformation. In the case of the antiferromagnetic Ising interaction,
when there is a geometrical frustration in the chain, the influence
of the on-site electron-electron attraction on the ground state, the
field and temperature dependences of the magnetization, and the
temperature dependences of the zero-field magnetic susceptibility and
the heat capacity are studied.

The phase diagram for the ground state is plotted in the plane $(\Delta
\tilde{I},\tilde{h})$. A modifications of this phase diagram induced
by the
transfer integral and the on-site attraction is represented in the form
of a
topological diagram $(\tilde{t},\tilde{U})$. It is shown that strengthening the attraction
along the \textquotedblleft equitopological\textquotedblright\ line
$\tilde{\mathcal{T}}(\tilde{t},\tilde{U})=\mathrm{const}$
is
reflected in the phase diagram ($\Delta\tilde{I},\tilde{h}$) as a
displacement
of the ground NAF state boundaries so that the corresponding confined area
becomes smaller.

As the on-site electron-electron attraction becomes stronger along the
\textquotedblleft equitopological\textquotedblright\ line, the
temperature
curves of the magnetization, at various fields, and the zero-field magnetic
susceptibility shift, as it was in the case of weakening the on-site electron-electron
repulsion \cite{lis2}. The temperature dependence of the
zero-field
heat capacity in a certain range $(\tilde{t},\tilde{U},\Delta\tilde{I})$
has
an additional maximum between the main and low-temperature maxima.

The results of this work also correspond to a simple
Ising--Hubbard chain, in which the nodes and the interstitial positions
are aligned, and the nodal Ising spin is coupled with those of the first and
second neighbors.

\vspace*{-5mm}
\rezume{%
Б.М.~Лісний}{АСИМЕТРИЧНИЙ РОМБІЧНИЙ ЛАНЦЮЖОК\\ ІЗІНГА--ГАББАРДА З
ПРИТЯГАННЯМ} {Розглянуто основний стан і термодинамічні властивості
асиметричного ромбічного ланцюжка Ізінга--Габбарда з одноцентровим
електрон-електронним притяганням, який є точно розв'язуваним за
допомогою де\-ко\-ра\-цій\-но-іте\-ра\-цій\-но\-го перетворення. У
випадку антиферомагнітної взаємодії Ізінга вивчено вплив цього
притягання на основний стан і температурну залежність
намагніченості, магнітної сприйнятливості та теплоємності.}

\end{document}